# Thermodynamic Behavior of a Perfect Axion Fluid with Negative Energy Density

Walter Christensen Jr.

Department of Physics
Cal State University, Fullerton
800 North State College Blvd. Fullerton, CA 92831
Email: wjchristense@csupomona.edu

February 09, 2006

**Abstract**: Starting with a perfect cosmological fluid represented by the energy momentum tensor $T_{\mu\nu}$, one class of frequency metrics that satisfies both Einstein's wave equation and the perfect fluid condition is given by: $g_{\mu\nu} = e^{i\omega t}\eta_{\mu\nu}$. Mathematically, such metrics indicate spacetime behaves locally like a simple harmonic oscillator. In the cosmological model presented here these small spacetime oscillations compress vacuum energy into a standing wave inside a dynamic Casimir cavity. At peak compression a phase shift occurs and the standing wave forms into a particle having relativistic mass-energy equal to the compressive work required to produce it. At this point the newly formed particle does isobaric work to expand the volume against the external pressure given by $T_{ii}$. Equilibrium is achieved when the collision rate on the volume's internal and external surface equalizes. By treating spacetime as a classical thermodynamic problem and oscillator, such quantities as the mass of the compressed particle--that of an axion, the radii of the initial and final volume of compression, and the angular frequency of compression, can be determined. During axion collision the photon frequency of the particle is calculated to be in the microwave range and inversely equal to the compression frequency needed to produce the particle. This suggests axion production is a source for the 2.7K cosmic background radiation and dark matter that pervades spacetime.

## I. Introduction

Within the study of general relativity perfect fluids are of cosmological importance[1]. In this paper the behavior of a perfect fluid with negative energy density is studied. Such a perfect fluid may be represented by the following energy momentum tensor:

$$T_{\mu\nu} = \frac{c^4}{8\pi G} \begin{bmatrix} -\frac{3\omega^2}{4} & 0 & 0 & 0 \\ 0 & \frac{\omega^2}{4} & 0 & 0 \\ 0 & 0 & \frac{\omega^2}{4} & 0 \\ 0 & 0 & 0 & \frac{\omega^2}{4} \end{bmatrix} \tag{1}$$

One class of frequency metrics[2] that satisfies both the Einstein's wave equation and the perfect fluid condition is given by:

$$g_{\mu\nu} = e^{i\omega t}\eta_{\mu\nu} \tag{2}$$

$\eta_{\mu\nu}$ is the flat Minkowski metric.

## II. Thermodynamic Model of Spacetime as a Simple Harmonic Oscillator

In the cosmological model presented in this paper, the energy moment tensor $T_{\mu\nu}$ suggests spacetime is filled with a perfect fluid having negative energy density. Being a perfect fluid it must be comprised of sparse, nearly massless particles that exists part and parcel with the vacuum energy pervading spacetime. The class of frequency metrics that satisfies both the perfect fluid condition and Einstein's general relativistic equation is given by $g_{\mu\nu} = e^{i\omega t}\eta_{\mu\nu}$. Mathematically, such metrics indicate spacetime behaves like simple harmonic oscillator. Furthermore if dynamic boundary conditions arise in spacetime, like that for Casimir effect, it is widely accepted particle creation is possible[3,4]. The cosmological model presented here relies on such a Casimir effect to create a dynamic cavity in which the vacuum energy is compressed into a standing wave. At peak compression a phase shift occurs and the standing wave forms into a particle having relativistic mass-energy equal to the compressive work required to produce it. Simultaneously, the newly formed particle does isobaric work to expand the volume against the constant external pressure given by $T_{ii}$. Equilibrium is achieved when the collision rate on the volume's internal and external surface equalizes. At this point the newly formed particle is released into the perfect fluid as dark matter.

Because the energy momentum tensor of equation (1) represents a perfect fluid of constant pressure and constant negative energy density it suggests the ideal gas law may be used to determine such quantities as the mass of the particle, the radii of the initial and final volumes and the angular frequency of compression and rarefaction. However, a slight modification to the gas law is expected in order to bring it in line with relativity.

This is necessary on two accounts. First, if the equation is to modified as simply as possible with relativistic attributes, in some way the moles (hence the mass of the perfect fluid) must be related to $E = mc^2$. Secondly, the number of moles for a dynamic universe has been shown to be a function of time[5,6] (i.e. of angular frequency).

At this juncture it must be questioned whether or not classical physics, general relativity, and quantum mechanics can be brought together to describe the cosmology presented here. The only valid answer is they can be only if each distinct description of nature is used piecewise, or separately, to ascertain various cosmological properties describing the universe. For example the energy momentum tensor of general relativity tells us the universe is filled with a perfect fluid. The associated metric tells us spacetime oscillates locally with angular frequency $\omega$. At this point we are basically done with general relativity but not with the fact the universe is filled with a perfect fluid that compresses and rarefies with angular frequency $\omega$. So whether it is the 19th century or 21st century, classical thermodynamics can be applied very effectively to describe this ideal gas pervading spacetime. Lastly, in regards to the Casimir effect, the historical path taken by Casimir is quite different than more current approaches[7]. Besides Casimir forces are a macroscopic phenomenon and have been used to explain particle production in dynamic spacetime processes[8,9]. . This is the approach taken in this paper to explain the forces that arise necessary to compress vacuum energy into a massive particle.

However, after all is said and done any proposed cosmological model must be verified through experimental evidence. If it turns out the model presented here can be validated experimentally then all references to classical physics must be removed in order to provide a more general theory for the cosmological theory presented. To verify this cosmological model experimentally, Casimir boundaries could be compressed in the separation distances calculated in this paper at multiple frequencies given by:

$$\frac{\omega}{2\pi n}, \quad \{n = 1,\ 2,\ 3...\} \tag{3}$$

### III. Visual Model and Calculations

To better visualize the Casimir forces that cause spacetime to oscillate[10] with frequency $\omega$, various figures are presented. Figure (1) shows the vacuum energy being compressed by Casimir forces into a standing wave inside a dynamic spherical cavity.

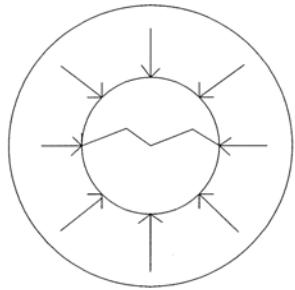

Fig. 1

Figure (2) shows microwaves radiating outwardly from the spherical surface. These microwaves are inversely equal to the spacetime frequency that accelerates the external vacuum energy, which is comprised of all fields, including electrodynamics ones.

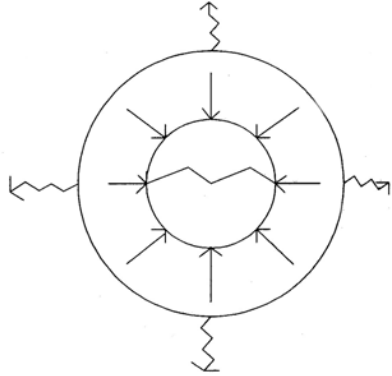

Fig. 2

After the phase shift occurs and the standing wave is compressed into a massive axion particle[11] equal to the isothermal work it took to produce it, the axion particle does isobaric work against the internal spherical boundary as shown in figure (3).

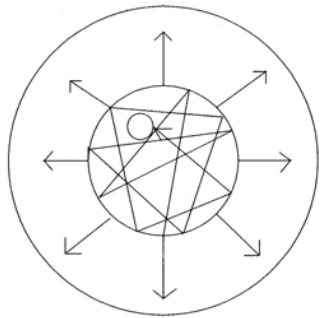

Fig. 3

An external constant pressure given by $T_{ii}$ is over come by the internal axion collision and expands the "hard boundary." This expansion continues until equilibrium is achieved. Equilibrium occurs when the collision rate of the internal and external surface of the volume becomes equal. When this occurs the axion particle is set free and contributes to cosmic dark matter[12]. This is a cosmic process is probably due to Le Chatellier's principle and could account for the distribution of the cosmic microwave radiation. Furthermore, Casimir forces can become attractive or repulsive depending on the shape and size of the boundaries[13].

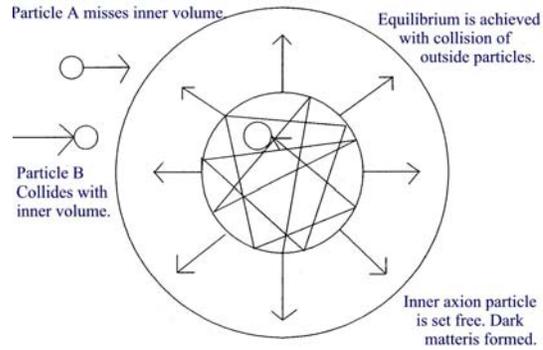

Fig. 4

The isobaric work to expand the volume is equal to the average kinetic energy of the external particles that stops the expansion[14]. The average kinetic energy of the external perfect fluid, or gas, is given by: $\overline{K} = \frac{3}{2}kT$ (k is the Boltzmann constant and T is the temperature of the external gas is assumed to be that of the cosmic background radiation of 2.7 K). The average kinetic energy is determined to be:

$$\overline{K} = 5.59 \times 10^{-23} \text{ Joules} \tag{4}$$

Isobaric work done by the internal axion particle is given by:

$$P\Delta V = nRT \tag{5}$$

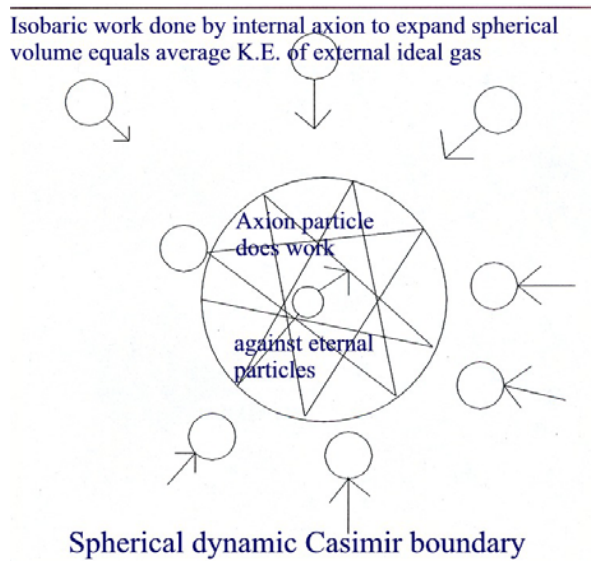

Figure (5)

This must isobaric work must be equal, or approximately equal, to the average kinetic energy $\overline{K}$ of external particles that stopped the expansion. Fortunately the temperature for single particle cancels and allows us to solve for the number of moles n:

$$n = \frac{3}{2}\left[\frac{k}{R}\right] = 2.49 \times 10^{-24} \text{ moles} \tag{6}$$

To determine the number of particles in this amount of moles, we multiply n by Avogadro's number $N_A$ to yield:

$$N \equiv nN_A = 1.5 \tag{7}$$

This value is very close to the predicted single axion particle produced per vacuum energy compression. The isothermal work of compression required to produce an axion at constant at 2.7K cosmic background temperature is given by: $PdV = \frac{nRT}{V}dV$. Integrating with respect to V obtains the work:

$$W = n_v RT \ln\left(\frac{V_f}{V_i}\right) \tag{8}$$

W is the work[15] in Joules. $n_v$ is the number of axion moles (rather a quasi-phase value of vacuum energy density) and is a function of frequency. R is the universal gas constant and T is the background radiation temperature in Kelvin. Assuming $n_v$ converges to n, the number of axion moles, at the boundary of phases, the natural log in equation (8) can now be solved for assuming isobaric and isothermal work are equivalent to the average kinetic energy of the eternal perfect fluid particles that stop volume expansion:

$$\ln\left(\frac{V_f}{V_i}\right) = \frac{\overline{K}}{nRT} = 1.00 \tag{9}$$

Solving for the ratio of the volumes gives the value:

$$\frac{V_f}{V_i} = e \tag{10}$$

Where the natural exponent e is the ratio volume of spacetime, in which the two volumes will be shown in section III to be related to spacetime frequency. In this natural ratio volume vacuum energy is compressed into an axion particle. It important inverse is given by:

$$\frac{V_i}{V_f} = \frac{1}{e} = 0.368 \tag{11}$$

It is interesting to note the natural log volume in equation (9) is very similar to the energy exchange in a collision between a free electron and heavy charge particle and the impact parameter natural log relationship of $\ln(\frac{r_{max}}{r_{min}})$ [16], where r is the impact parameter ($r_{min} < r < r_{max}$) and related to the radius in change in isobaric volume. If an axion particle is converted into a photon during such a collision it must have a frequency[17] related to the frequency $\frac{\omega}{2\pi}$ of spacetime compression that forms standing wave. The frequency of the axion particle is computed to be in the range of the microwave background radiation, and is assumed to be one source for the 2.7K cosmic background radiation. It must also be equivalent to the relativistic mass-energy of the particle. Einstein's equation states $E = mc^2$ and by hypothesis must be equal to the work that produced it, and thus to the kinetic energy of the particle and its mass. The mass is of this particle is computed to be:

$$m_a = \frac{\overline{K}}{(c^2)} = 6.25 \times 10^{-40} \text{ Kg} \qquad (12)$$

Or $3.5 \times 10^{-4} \text{ eV}$, and is in strong agreement with the mass of an axion particle[18,19] currently sought Lawrence Livermore National Laboratory, the University of California at Berkeley (UCB), the University of Florida (UF), and the National Radio Astronomy Observatory (NRAO)[20] Also note the average kinetic energy divided by the Bohr magneton equals 6.0 Tesla.

**IV. Energy Momentum Tensor Applied to the Ideal Gas Law**

In this section the value for the angular frequency of spacetime $\omega$ is computed. The constant external pressure given by the energy momentum tensor of equation (1) implies isobaric work. The component of the pressure is given by:

$$T_{ii} = \frac{\omega^2 c^4}{4 \cdot (8\pi G)}, \quad (i = 1,2,3) \qquad (13)$$

$G = 6.67 \times 10^{-11} \frac{N \cdot m^2}{kg^2}$, $c = 2.99 \times 10^8 \frac{m}{s}$. The magnitude of the pressure is computed from its components given by $T_{\mu\nu}$:

$$|P| = \frac{\sqrt{3}\omega^2 c^4}{4 \cdot (8\pi G)} \qquad (14)$$

$\omega$ is then computed from the ideal gas law and isobaric work:

$$|P| = \frac{\sqrt{3}\omega^2 c^4}{4\cdot(8\pi G)} = \frac{nRT}{\Delta V} \tag{15}$$

However at this point a slight adjustment must be made to bring the moles and volume in line with relativity, and to do so as simply as possible. Examining the left hand side of equation (15), and knowing $(8\pi G)$ is fundamental to general relativity, the simplest choice for the volume is to assign it the value of $8\pi\overline{G}\ m^3$, where $\overline{G}$ is defined to be the unit-less value of $\overline{G} = 6.67 \times 10^{-11}$. This is the volume of expansion. The number of moles, n, (hence the mass of the perfect fluid particles) must be related to $E = mc^2$. Thus $n \to \overline{c}^2 n$, where $\overline{c} \equiv 2.99 \times 10^8$ is without units. This allows c and G to be set equal to one and the same values will result for angular frequency. Equation (15) becomes:

$$\frac{\sqrt{3}\omega^2 c^4}{4\cdot(8\pi G)} = \frac{\overline{c}^2 nRT}{(8\pi\overline{G})} \tag{16}$$

Its choice will be substantiated by calculation in section III. Such a volume designation suggests axion particle production might give rise to the gravitational constant G. Where:

$$\Delta V = 8\pi\overline{G}\ m^3 = 2.0 \times 10^{-9}\ m^3 \tag{17}$$

The natural angular frequency $\omega$ squared, has the relativistic form $(ct)^{-2}$. However the numerator in equation (15) has $c^4$. The time part of $\omega$, called $\omega'$, is calculated to be:

$$\omega' = 1.14 \times 10^{-11}\ \sec^{-1} \tag{18}$$

The energy density $T_{00}$ can now be computed:

$$T_{00} = -\frac{3\cdot c^4}{4\cdot(8\pi G)}\omega^2 = -\frac{3\cdot c^2 (1.14 \times 10^{-11})^2}{4\cdot(8\pi G)} = -4.36 \times 10^3\ \frac{J}{m^3} \tag{19}$$

The physical meaning of negative sign for energy density can be understood because the energy density being considered is inside the small volume $8\pi\overline{G}\ m^3$ as can be discerned from equation (19) before compression. To show this, the energy is divided by $c^2$ yields: $-4.87 \times 10^{-14}\ \frac{kg}{m^3}$. The equivalent mass in that volume is given by: $9.74 \times 10^{-23}$ kg, which far exceeds the axion particle mass of $m_a = 6.25 \times 10^{-40}$ kg. This implies only a small amount of the vacuum energy is compressed into an axion particle. The metric tells us spacetime is dynamic. The energy momentum tensor tells us pressure and negative energy density is constant. This is almost a contradiction except if the pressure considered is outside the volume $8\pi\overline{G}\ m^3$ to which work must be done against. The energy density is internal to which a particle is compressed. As long as the volume is intact that amount of energy density remains constant. In actually energy is neither created nor destroyed in the process only converted from vacuum energy.

**V. Spacetime Interval**

The spacetime interval with the metric from equation (2) is explored.

$$ds^2 = e^{i\omega t}\eta_{\mu\nu}dx^\mu dx^\nu = e^{i\omega t}(-dt^2 + dx^2 + dy^2 + dz^2) \qquad (20)$$

Taking the real part of the interval (with angular frequency $\omega = 1.14 \times 10^{-11} \sec^{-1}$) leads to:

$$ds^2 = \cos(1.14 \times 10^{-11} t)(-dt^2 + dx^2 + dy^2 + dz^2) \qquad (21)$$

The period of space-time compression is given by $\frac{2\pi}{\omega}$. Thus:

$$T = 5.51 \times 10^{11} \sec \qquad (22)$$

The frequency of spacetime compression is given by:

$$\nu = 1.81 \times 10^{-12} \sec^{-1} \qquad (23)$$

Furthermore, the work to compress vacuum energy into mass should be directly related to the photon frequency emitted during collision of axion particles. By assumption the photon should have a frequency of that of the background microwave radiation. By examination the frequency of spacetime compression it is inversely equal to the frequency of compression. However, as the compression maximizes, the cosine function representing compression will be at a minimum. The reciprocal is true for the emission of the photon. Therefore the axion radiation frequency is inversely equal to the spacetime frequency.

$$\nu_{2.7} = 5.51 \times 10^{11} \sec^{-1} \qquad (24)$$

This frequency is within the range of microwave radiation frequency. The energy associated with this frequency is given by:

$$E = \hbar\nu_{2.7} = 5.79 \times 10^{-23} \text{ Joules} \qquad (25)$$

Comparing this axion collision energy to the compression energy of the axion particle yields a 3.5% error difference.

## VI. Quantum approximation

Because the axion particle is compressed from the vacuum energy, axion particle production must be related to the Casimir effect--an outcome of quantum field theory, in which changes in the vacuum energy can be attractive or repulsive. Canonically, a field at each point in space under second Quantization is considered to be a simple harmonic oscillator. This harmonic oscillation is also suggested for spacetime as well by the metric given by equation (2). The Casimir effect assumes standing waves of the electromagnetic field in the cavity. Likewise the axion particle field is assumed to form a standing wave during spacetime compression. This implies the spacetime volume of compression (not the ratio of volumes) can be computed directly from the Casimir effect result[21,22,23]:

$$\frac{\langle E \rangle}{A} = \frac{\hbar c \pi^2}{6a^3} \varsigma(-3) \tag{26}$$

$\langle E \rangle$ is the vacuum energy expectation value, A is the area of the metal plates in the Casimir's calculation, a is the distance between those metal plates. The zeta function is typically given a value of: $\varsigma(-3) = \frac{-1}{120}$ As an approximation, between a classical and quantum effect, we set the zeta value to one because the expectation value for the vacuum energy density is, under a classical approximation held to be the average kinetic energy, the same energy from which the axion particle was compressed. Furthermore, for standing sound wave to occur, the length for which compression takes place must be equal to multiple wavelengths. The wavelength of the axion particle was computed to be in microwave length range. Thus, as an approximation, the compression volume is expected to have microwave length. To compute this volume a gaussian surface of volume $a^3$ is formed around the volume of compression. The area simply becomes $a^2$. Rearranging equation (26) for a yields:

$$a = \frac{\hbar c \pi^2}{6 \langle E \rangle} \tag{27}$$

Substituting $\langle E \rangle = \overline{K}$ yields the value:

$$a_{cube} = 9.24 \times 10^{-4} \text{ meters} \tag{28}$$

The Casmir energy for the massless scalar field inside a sphere of radius a in dimension 3 with Dirichlet boundary conditions is given by[24]:

$$E_{Cm} = \frac{\hbar c}{a} \left( 0.0044 + \frac{1}{630\pi} \left[ \frac{1}{s} + \ln\left(\frac{\mu c a}{\hbar}\right)^2 \right] \right) \tag{29}$$

Here $\mu$ is the axion mass and "a" is the radius of the sphere. The natural log term simply vanishes because $\frac{\mu c a}{\hbar}$ is equal to unity (or can be made so with a miniscule adjustment to the axion mass $\mu$. In general the second and third term cancels, which implies $\frac{1}{s} \to -630\pi$. Keeping only the first term, and using the axion energy for the Casimir energy, yields a value for a:

$$a_{sphere} = 5.42 \times 10^{-4} \text{ meters} \tag{30}$$

The length of a, is in the microwave range and supports the calculation previously for $\omega$. The axion massive particle velocity is ($v = c\sqrt{2}$). However in the present the case considered is instantaneously before full compression. The particle is still in wave form and cannot travel faster than c. The wavelength is thus computed to be:

$\lambda = \frac{c}{v_{2.7}} = 5.42 \times 10^{-4}$ meters, which is precisely the value computed for radius of the sphere. This implies inside the sphere fits two wavelengths, meaning particle parity. The volume of for the cube of compression is computed to be:

$$a^3 = 7.88 \times 10^{-10} \text{ m}^3 \tag{31}$$

Since $a^3$ is at full compression it is equal to $V_i$. With this information we can check the result in equation (9) independently. The ratio of $V_i$ to $V_f$ is compute to be:

$$\frac{a^3}{\Delta V} = 0.394 \approx \frac{1}{e} \tag{32}$$

where $\Delta V = 8\pi \overline{G} \text{ m}^3 = 2.0 \times 10^{-9} \text{ m}^3$. This is the value we would expect from equation (11). Notice how wonderful this result reinforces the choice for the gravitational volume given by equation (17). By comparing equation (32) with equation (11), it is realized $V_f = \Delta V$. That is:

$$V_f = e \cdot a^3 = 2.0 \times 10^{-9} \text{ m}^3 \tag{33}$$

Comparing the final and initial volumes yields a radius of compression of about 0.75 mm. The number of moles in $a^3$ can be computed from the ideal gas law and General relativistic pressure, rearranging yields: $n\overline{c}^2 = \frac{\sqrt{3}\omega^2 c^4}{4 \cdot (8\pi GRT)} V_f$. The number of moles (keeping in mind $\omega$ has inverse $c^2$) is computed to be:

$$n = 2.99 \times 10^{-24} \tag{34}$$

The number of axion particles in volume $a^3$ is n times Avogadro's number $6.02 \times 10^{23}$ and equals: 1.8. This number is close to unity as expected during compression to produce a single particle. Each particle was computed to have a mass of $6.25 \times 10^{-40}$ kg. Assuming one particle and dividing this by the volume yields and energy density of:

$$\rho = 3.13 \times 10^{-31} \frac{\text{kg}}{\text{m}^3} \qquad (35)$$

This density falls well beneath the experimental upper limit for axion density dark matter.

## VII. Conclusion

Starting from a particular energy momentum tensor representing a perfect fluid, one class of frequency metrics that satisfies both the wave equation of General Relativity and the perfect fluid condition is: $g_{\mu\nu} = e^{i\omega t} \eta_{\mu\nu}$. Thermodynamic behavior of the perfect fluid is treated classically and applied to the cosmic background radiation. The mass of the particles comprising this perfect fluid is determined to be that of an axion particle. The metric also imposes the condition spacetime is locally a simple harmonic oscillator. The frequency of spacetime is shown to be inversely equal to the microwave photon frequency for colliding axion particles. It is concluded axion production emits the 2.7K microwave background radiation during compression as well as collision. The photon frequency associated with axion is inversely related to the angular frequency of spacetime. During axion collision its photon frequency is equal to the cosmic microwave radiation. Experimentally this could be verified if microwave radiation and axion particle from vacuum energy could be created from dynamic Casimir boundaries compressing together and rarefying apart at multiple frequencies given by:

$$\omega \cdot n = 1.14 \times 10^{-11} \sec^{-1} \cdot n \qquad (36)$$

where $n = \{1, 2, 3...\}$.